\documentclass[10pt,a4paper]{article}
\makeatletter
\usepackage[dvipsnames]{xcolor}
\usepackage{lipsum}
\usepackage{tipa}
\usepackage{tikz}
\usepackage{tikz-cd}
\usetikzlibrary{matrix,arrows,decorations.pathmorphing}
\newcounter{magicrownumbers}

\usepackage[pdftex,bookmarks=true,bookmarksnumbered=true]{hyperref}
\hypersetup{linkbordercolor=ForestGreen!25, citebordercolor=Goldenrod!60}
\usepackage{graphicx}
\usepackage{stmaryrd}
\usepackage{twoopt}
\usepackage{amssymb}
\usepackage{amsmath}
\usepackage{amsthm}
\usepackage{mathabx}
\usepackage{amsfonts}
\usepackage{amscd}
\usepackage{url}
\DeclareMathAlphabet{\mathpzc}{OT1}{pzc}{m}{it}
\usepackage[all]{xy}

\newcommand{\hil}{\mathcal{H}}

\newcommand{\cl}{\mathrm{cl}}

\newcommand{\5}{\hspace{0,5cm}}
\newcommand{\3}{\vspace{0,3cm}}

\newcommand{\tx}[1]{\mathpzc{#1}}
\newcommand{\fp}[2]{\mspace{1mu} {}_{#1} \mspace{-6mu} \times_{#2}}

\newcommand{\gpdact}[4]{\begin{tikzpicture}
\path (2,1) node (a) {${#1}$} (2,0) node (c) {${#3}$} (0,0) node (b) {${#2}$};
\begin{scope}
\draw[->] (0,0.4) arc (180:90:44pt and 18pt);
\draw[->] (a) -- node [right] {${#4}$} (c) ;
\draw[->] (0.4,0.05) -- (1.6,0.05) ;
\draw[->] (0.4,-0.05) -- (1.6,-0.05) ;
\end{scope}
\end{tikzpicture}}

\title{Spectral Triples on Proper {\' E}tale Groupoids}
\date{}
\author{Antti J. Harju\footnote{Copernicus Center for Interdisciplinary Studies, Krakow, Poland} \footnote{harjuaj@gmail.com}}
\begin{document}
\maketitle

\begin{abstract}
A proper {\' e}tale Lie groupoid is modelled as a (noncommutative) spectral geometric space. The spectral triple is built on the algebra of smooth functions on the groupoid base which are invariant under the groupoid action.  Stiefel-Whitney classes in Lie groupoid cohomology are introduced to measure the orientability of the tangent bundle and the obstruction to lift the tangent bundle to a spinor bundle. In the case of an orientable and spin Lie groupoid, an invariant spinor bundle and an invariant Dirac operator will be constructed. This data gives rise to a spectral triple. The algebraic orientability axiom in noncommutative geometry is reformulated to make it compatible with the geometric model.\3


\end{abstract}

\section*{Introduction}

The goal of this work is to take a step towards a unification of Lie groupoid theory and (noncommutative) spectral geometry. We will restrict the study to the proper {\' e}tale Lie groupoids. The classical orbifolds are represented by a Lie groupoid with these properties \cite{MPro97}, \cite{Moe02}. Roughly speaking, orbifold theory is a geometric model for manifolds subject to local smooth actions of finite group \cite{Sat56}. The essential feature in this formalism is that one can have several different groups acting on different regions of the manifolds. The Lie groupoid theory puts the orbifold theory to a more general geometric context. On the other hand, one can understand Lie groupoids as "local charts" of differentiable stacks. The dictionary between these two theories is provided in \cite{BX11}. In this work, the relationship to yet another geometric theory, spectral geometry, will be studied. The approach in the spectral geometry is to put the geometric structures in operator theoretic framework: in the case of compact manifolds, one can recover the whole riemannian geometry from such data, \cite{Con13}. This theory is often referred to noncommutative geometry since it allows to study noncommutative algebras as well, \cite{Con94}. In fact, some interesting noncommutative deformations of function algebras on groupoids have been developed recently, see for example \cite{BF12}. However, the geometric study of these algebras (in terms of spectral triples) still lacks the classical limit. 

The spectral geometric model is based on a topological algebra. In this work, the algebras of interest are the subalgebras of smooth functions on the base of the groupoids which are invariant under the action of the groupoid arrows. One also needs a Dirac operator and a complex Hilbert space of spinors. The role of the Dirac operator is to measure the metric properties. The spinor space has a natural action by the groupoid arrows which is inherited from the action of the groupoid on the tangent bundle by the local diffeomorphism associated to the arrows. The spectral triple model will be based on the invariance under the action. 

The approach in this work builds on a localization of a proper {\' e}tale Lie groupoid. If the base of a Lie groupoid is given a cover, then one can decompose the manifolds of arrows to components consisting of the arrows between these cover sheets. This defines a localization of the whole groupoid structure. We shall use the local structure for the cohomological classification of vector bundles and for the lifting and reduction problems of their structure groups. These problems arise in the study of orientability and spin property on proper {\' e}tale Lie groupoids. The localization strategy has been applied elsewhere in the literature, for example in the theory of gerbes and twisted K-theory on orbifold groupoids, \cite{LU04b}, \cite{TX06}. 

In the differential geometric context, one can build spinor bundles as follows. The tangent bundle $\tau X$ on a manifold $X$ is equipped with a riemannian structure which gives a reduction for the structure group of $\tau X$ to the orthogonal group $O_n$. Provided that $\tau X$ is orientable, the structure group reduces further to $SO_n$. The obstruction for this reduction is measured by nontriviality of the first Stiefel-Whitney class $w_1(\tau X) \in H^1(X, \mathbb{Z}_2)$. The bundle structure and the Stiefel-Whitney class are convenient to code in a {\v C}ech cohomology data associated to a fixed good cover. There is a covering homomorphism $\varpi: \text{Spin}_n \rightarrow SO_n$ and if $\tau X$ is orientable, one can try to lift the transition cocycle of the bundle $\tau X$ through $\varpi$. The obstruction for this lift is measured by nontriviality of the second Stiefel-Whitney class $w_2(\tau X) \in H^2(X, \mathbb{Z}_2)$ which is convenient to realize in the {\v C}ech cohomology as well. If the obstruction vanishes one gets a spinor bundle by a reconstruction from the transition cocycle. In addition, one gets a Dirac operator acting on the spinors and after a Hilbert space completion, there is a classical spectral triple if $X$ is geodesically complete. 

In the case of a proper {\' e}tale Lie groupoid $\Theta \rightrightarrows X$, there is a tangent bundle $\tau X$ on the base manifold $X$. It is naturally equipped with an action of the groupoid $\Theta$ by the local diffeomorphisms associated to the arrows $\Theta$. We shall proceed by localizing the groupoid and writing down a transition cocycle for $\tau X$ which determines a class in the groupoid cohomology. The cocycle contains the information of the topological structure of $\tau X$ and of the $\Theta$-action. The bundle $\tau X$ can be equipped with a riemannian structure for which the groupoid acts orthogonally. This leads to an $O_n$ reduction of the transition cocycle. The $SO_n$ reduction is associated to the orientability of the groupoid. A Stiefel-Whitney class is introduced to measure the obstruction to be able to do the reduction. Then $\text{Spin}_n$ lifting of the $SO_n$ valued cocycles will be studied: this leads to the spin structures and to the second Stiefel-Whitney class in the groupoid cohomology. A spinor bundle with a groupoid action is then reconstructed from the transition cocycles. The space of sections has a $\Theta$-invariant subspace which is completed to a Hilbert space. The Dirac operator in this model is also $\Theta$-invariant and therefore acts on the invariant subspace. This will lead to a spectral triple for the algebra $C^{\infty}(X)^{\Theta}$ - the $\Theta$-invariant subalgebra of the algebra of smooth functions. 

There is an approach to define spinors on Lie groupoids which applies the theory of Lie algebroids, \cite{PPT13}. More specifically, the tangent bundle in this formalism
 is defined to be the kernel subbundle for the tangent map $dt$ of the target morphism $t$ pulled back to the base by the unit morphism in the Lie groupoid. In the case under consideration $t$ is a local diffeomorphism and therefore the tangent spaces would be the zero vector spaces. Therefore this approach would not be useful here.

The structures of spectral triples associated to orbifolds have been studied in \cite{RV08}. However, in this reference the consideration has been restricted to the global action orbifolds which are manifolds subject to a global action by a finite group. It is pointed out in \cite{RV08} that in this special case the invariant spectral triple does not satisfy the algebraic orientability axiom for a spectral triple \cite{Con13}. We shall reformulate the algebraic orientation so that the new orientation captures the  geometric orientation of a Lie groupoid. However, the invariant spectral triples do not carry a sufficient amount of data for one to be able to write down the new orientation in terms of the invariant structures involved. Therefore an extension for the spectral geometric model is needed. To this end, we develop an explicit model for a spectral triple on the smooth convolution algebra $C^{\infty}_c(\Theta)$ of an effective proper {\' e}tale Lie groupoid. Similar model has been studied in \cite{GL13}. \3 

I wish to thank Andrzej Sitarz and the referee for valuable comments leading to several improvements. \3

\noindent \textbf{Notation.} We use the symbol $X_{\bullet}$ to denote a Lie groupoid with a base manifold $X_{(0)}$ and a manifold of $k$-times composable arrows $X_{(k)}$ for all $k \in \mathbb{N}$. We also write occasionally $X_{(1)} = \Theta$ and $X_{(0)} = X$ and then use the symbol $\Theta \rightrightarrows X$ to denote the groupoid. $\Theta$ and $X$ are locally compact smooth manifolds which are both Hausdorff and second countable. The target and source maps: $s,t : \Theta \rightarrow X$ are smooth submersions. $X_{\bullet}$ has a simplicial structure. The face maps $\partial^i_k: X_{(k)} \rightarrow X_{(k-1)}$ are defined by
\begin{eqnarray}\label{arrow}
\partial^i_k(\sigma_1, \ldots, \sigma_k) =  \left\{ \begin{array}{ll} (\sigma_{2}, \ldots, \sigma_k), & i = 0 \\
(\sigma_1, \ldots, \sigma_{i-1}, \sigma_{i} \sigma_{i+1}, \sigma_{i+2}, \ldots, \sigma_k), & i \neq 0,k \\ \noindent
(\sigma_1, \ldots, \sigma_{k-1}), & i = k \end{array} \right. \noindent
\end{eqnarray}
and the degeneracy maps $s^i_k: X_{(k)} \rightarrow X_{(k+1)}$ are defined by 
\begin{eqnarray*}
s^i_k(\sigma_1, \ldots, \sigma_k) = (\sigma_1, \ldots, \sigma_{i-1}, \textbf{1}_{t(\sigma_i)}, \sigma_{i},  \ldots, \sigma_k). 
\end{eqnarray*}
$\textbf{1}_{x}$ denotes the unit morphism at $x \in X$ on the base. If $U$ and $V$ are subsets in $X$ we define the following subspaces in $\Theta$: 
\begin{eqnarray*}
\Theta_U = s^{-1}(U), \5 \Theta^V = t^{-1}(V) \5 \text{and}\5 \Theta_U^V = s^{-1}(U) \cap t^{-1}(V)
\end{eqnarray*}
If $x \in X$ and $U = V = \{x\}$ then $\Theta_x^x$ is the isotropy group at $x$. The orbit space (coarse moduli space) of $X_{\bullet}$ is denoted by $|\Theta|$.

A groupoid $X_{\bullet}$ is proper if $(s,t): \Theta \rightarrow X \times X$ is a proper map, and {\' e}tale if $s$ and $t$ are local diffeomorphisms.  All groupoids are Lie groupoids in this work. The parameter $n$ will denote the dimension of the base manifold $X$ everywhere below.
 
\section{Local Structure}

\noindent \textbf{1.1.} Let $X_{\bullet}$ be a Lie groupoid. For each arrow $\sigma \in \Theta_x$ there is an open neighborhood $U$ of $x$ and a local section of $s$, $\hat{\sigma}: U \rightarrow \Theta$, such that $\hat{\sigma}(x) = \sigma$ and $t \circ \hat{\sigma}$ is an open embedding. The map $\hat{\sigma}$ is called a local bisection of $\sigma$. In the case of an {\' e}tale Lie groupoid, any two local bisections of an arrow agree on their common domain. Now the composition
\begin{eqnarray*}
 \varphi_{\sigma} = t \circ \hat{\sigma}: U \rightarrow \varphi_{\sigma}(U). 
\end{eqnarray*}
defines a local diffeomorphism. Denote by $\Delta(\Theta)$ the set of germs of the local diffeomorphism arising from the local bisections.  An {\' e}tale Lie groupoid is defined to be effective if the map $\Theta \rightarrow \Delta(\Theta)$ which sends the arrow $\sigma$ to the germ of the local diffeomorphism $\varphi_{\sigma}$ is injective. 

Suppose that $X_{\bullet}$ is a proper {\' e}tale Lie groupoid. For all $x \in X$ there is an open neighborhood $U$ of $x$ such that $X_{\bullet}$ localizes to a transformation groupoid $\Theta^x_x \ltimes U \rightrightarrows U$ where $\Theta_x^x$ is the isotropy group at $x$. The isotropy group acts on $U$ through the local diffeomorphisms associated to the local bisections of the arrows in $\Theta_x^x$. The open set $U$ can be taken to be an arbitrarily small euclidean ball so that the group $\Theta_x^x$ acts on it as a finite subgroup in $GL_n$. The orbit space $|\Theta|$ can be given a structure of an orbifold so that the orbifold charts are determined by these local action groupoids and by the natural groupoid projection onto the orbit space, \cite{MPro97}, \cite{MM03}. 

Consider an open cover $\{N_{a} : a \in I\}$ of the groupoid base $X$. Since we are working with manifolds the index set $I$ can be taken to be numerable. The {\v C}ech groupoid $\check{X}_{\bullet}$ associated to the open cover is defined by 
\begin{eqnarray*}
\coprod_{ab} \Theta_{N_b}^{N_a} \rightrightarrows \coprod_a N_a, 
\end{eqnarray*}
where 
\begin{eqnarray*}
\Theta_{N_b}^{N_a} = \{\sigma \in \Theta: s(\sigma) \in N_b, t(\sigma) \in N_a \}. 
\end{eqnarray*}
and the source and the target maps are the restrictions of the source and target maps in $\Theta$. The {\v C}ech groupoid $\check{X}_{\bullet}$ is always Morita equivalent to $X_{\bullet}$. In particular, the Lie groupoid cohomology can be computed by passing to a suitable {\v C}ech groupoid. Since we are working with proper {\' e}tale groupoids, we can always choose an open cover for $X$ so that the groupoid localizes as a transformation groupoid over each of its component. Since the components can be chosen to be arbitrarily small euclidean balls, the cover can be assumed to be good. For the same reason we can assume that each $N_a$ is equipped with its own coordinate functions: $\varphi_a: N_{a} \rightarrow \mathbb{R}^n$. A good coordinate cover means a cover which is good and its components are equipped with coordinate functions. \3

\noindent \textbf{1.2.}  A vector bundle over a Lie groupoid $X_{\bullet}$ is a smooth vector bundle on the base with a typical fibre $V$ which is equipped with a $\Theta$-action
\begin{center}
\gpdact{\xi}{\Theta}{X}{\pi}
\end{center}
The domain of the action is the fibre product $\Theta \fp{s}{\pi} \xi$  and if $\sigma \in \Theta_x$ is an arrow and $u$ is a vector in the fibre $\xi_x$ over $x \in X$, then $\sigma$ acts by
\begin{eqnarray*}
(\sigma, u_x) \mapsto (\rho(\sigma) u)_{\sigma x} 
\end{eqnarray*}
so that $u_x$ maps to the fibre $\xi_{\sigma x}$ over $\sigma x \in X$ and $\rho: \Theta \rightarrow GL(V)$ is required to satisfy
\begin{eqnarray*}
\rho(\tau) \rho(\sigma) = \rho(\tau \sigma), \5 \rho(\textbf{1}_x) = \iota 
\end{eqnarray*}
for all $(\tau, \sigma) \in X_{(2)}$ and unit arrows $\textbf{1}_x$. Here the vector space $V$ is always taken to be of finite rank but one can take it to be real or complex. 

An inner product in a vector bundle $\xi$ over the groupoid $X_{\bullet}$ is a smoothly varying inner product in the fibres of $\xi$ which is invariant under the action of $\Theta$: 
\begin{eqnarray*}
(\rho(\sigma) u_1, \rho(\sigma) u_2)_{t(\sigma)} = (u_1, u_2)_{s(\sigma)}
\end{eqnarray*}
for all $\sigma \in \Theta$ and vectors $u_1,u_2$ in the fibre over $s(\sigma)$. One can use the existence of an ordinary inner product in the bundle $\xi$ together with an averaging trick over the $s$-fibres of $X_{\bullet}$ to construct a $\Theta$-invariant inner product in $\xi$. This can be done by applying a Haar system and a cutoff. A right Haar system is a collection of measures $\mu = \{\mu_x\}_{x \in X}$ which are supported in $\Theta_x = s^{-1}(x)$ such that:
\begin{quote}
\textbf{1.} For all $f \in C^{\infty}_c(\Theta)$, $x \mapsto \int_{\sigma \in \Theta_x} f(\sigma) \mu_x(d \sigma)$ is smooth function on $X$. 

\textbf{2.} The measures are invariant under the right translations by the arrows: 
\begin{eqnarray*}
\int_{\sigma \in \Theta_{t(\sigma')}} f(\sigma \sigma') \mu_{t(\sigma')}(d \sigma) = \int_{\sigma \in \Theta_{s(\sigma')}} f(\sigma) \mu_{s(\sigma')} (d \sigma).
\end{eqnarray*}
\end{quote}
A function $c: X \rightarrow \mathbb{R}_+$ is called a cutoff if the integration of $c \circ t$ over each $s$-fibre on $X$ satisfies
\begin{eqnarray*}
\int_{\sigma \in \Theta_x} c(t(\sigma)) \mu_x(d \sigma) = 1
\end{eqnarray*}
and for all compact sets $K \subset X$, the support of $(c \circ t)|_{\Theta_K}$ is compact. A Haar measure and a cutoff always exists in a proper Lie groupoid. Now a $\Theta$-invariant inner product can be defined by, \cite{PPT13b}: 
\begin{eqnarray*}
(v,w)^I_x = \int_{\tau \in \Theta_x} c(t(\tau)) (\rho(\tau)v, \rho(\tau)w)_{t(\tau)} \mu_x(d \tau)
\end{eqnarray*}
if $(\cdot, \cdot)$ is any ordinary smooth inner product in the bundle $\xi$. In conclusion. \3

\noindent \textbf{Proposition 1.} A vector bundle over a proper {\' e}tale groupoid can be equipped with an inner product. \3

\noindent \textbf{1.3.} Let $\tau X$ denote the tangent bundle over the base $X$. We can apply the elements of $\Delta(\Theta)$ to define a $\Theta$-action on $\tau X$. For any arrow $\sigma: x \rightarrow y$ there is a germ of local diffeomorphisms $\varphi_{\sigma}$ which amounts to define the differential map $(d \varphi_{\sigma})_x$. This gives the action 
\begin{eqnarray*}
\Theta \fp{s}{\pi} \tau X \ni (\sigma, [x,v]) \mapsto [\sigma x, (d \varphi_{\sigma})_x(v)] \in \tau X,
\end{eqnarray*}
where we have applied a local trivialization of $\tau X$ around $x$ and $\sigma x$ in the standard way. This is indeed well defined since for a composable pair $(\tau, \sigma) \in X_{(2)}$ the differentials satisfy 
\begin{eqnarray*}
(d \varphi_{\tau})_{t(\sigma)} (d \varphi_{\sigma})_{s(\sigma)} = d(\varphi_{\tau} \circ \varphi_{\sigma})_{s(\sigma)} = d(\varphi_{\tau \circ \sigma})_{s(\sigma)}.
\end{eqnarray*}
The last equality holds since the local diffeomorphisms arising from the arrows are uniquely determined in a neighborhood of $s(\sigma)$. Proposition 1 provides a $\Theta$-invariant real inner product for the bundle $\tau X$. A groupoid equipped with an inner product in $\tau X$ will be referred to a riemannian groupoid. 

A riemannian groupoid is defined to be oriented if the base $X$ has a coordinate system such that, for any arrow $\sigma \in \Theta$, the Jacobian matrix associated to the linear transformation  
\begin{eqnarray*}
(d \varphi_{\sigma})_x: \tau X_x \rightarrow \tau X_{\sigma x}
\end{eqnarray*}
is an element in the group of invertible linear transformations with positive determinant $GL^+_n$. Notice that this implies that $\tau X$ is an oriented vector bundle in the ordinary sense: at each $x \in X$ there is the identity morphism $\textbf{1}_x$ and the local diffeomorphisms associated to the unit arrows are identity maps. Especially the differentials of the unit maps are simply the Jacobians associated to the coordinate transformations and in the case of an orientable groupoid it has a $GL^+_n$ reduction implying that $\tau X$ is an orientable in the ordinary sense. \3

\noindent \textbf{1.4.} The goal in the following is to describe the tangent bundle and the orientability in terms of cohomology theoretic data. For this we shall apply the sheaf cohomology theory on Lie groupoids, \cite{BX11}. This theory is invariant under the Morita equivalences. The general principle is to work with a double complex in which one of the directions is determined by a cocycle complex arising from the simplicial structure of the arrows and another direction is determined by an injective resolution.  In the case of a Lie groupoid, the injective resolutions can be replaced by {\v C}ech resolutions, up to an isomorphism.

Let $X_{\bullet}$ be a Lie groupoid. For all $k > 1$ we have the $k+1$ arrows 
\begin{eqnarray*}
\partial^i_{k}: X_{(k)} \rightarrow X_{(k-1)}
\end{eqnarray*}
associated to the simplicial structure, \eqref{arrow}. Each $X_{(k)}$ is a smooth manifold and we can choose good open covers $\mathfrak{N}_{(k)} = \{N_{(k),i}: i \in I_{(k)}\}$ for each $X_{(k)}$. The index sets $I_{(k)}$ can be chosen to be countable. Let $\tx{G}$ denote a sheaf of smooth functions valued in an abelian group $G$. The sheaf cohomology groups $\check{H}^*(X_{\bullet}, \tx{G}_{\bullet})$ can be computed from the double complex 
\begin{center}
\begin{tikzpicture}
\path (0,2) node (u1) {$\cdots$} (3,2) node (u2) {$\cdots$} (6,2) node (u3) {$\cdots$}
      (0,1) node (d1) {$\check{C}^1(\mathfrak{N}_{(0)},\tx{G})$} (3,1) node (d2) {$\check{C}^1(\mathfrak{N}_{(1)},\tx{G})$} (6,1) node (d3) {$\check{C}^1(\mathfrak{N}_{(2)},\tx{G})$}  (8,1) node (d4)  {$\cdots$}
			(0,0) node (dd1) {$\check{C}^0(\mathfrak{N}_{(0)},\tx{G})$}  (3,0) node (dd2) {$\check{C}^0(\mathfrak{N}_{(1)},\tx{G})$}  (6,0) node (dd3) {$\check{C}^0(\mathfrak{N}_{(2)},\tx{G})$}  (8,0) node (dd4)  {$\cdots$};
\begin{scope}
\draw[->] (d1) -- node [above] {$\partial^*_0$} (d2); \draw[->] (d2) -- node [above] {$\partial^*_1$} (d3); \draw[->] (d3) -- node [above] {$\partial^*_2$} (d4); 
\draw[->] (d2) --   node [right] {$\delta_1$} (u2); \draw[->] (d3) --  node [right] {$\delta_1$}  (u3); \draw[->] (d1) -- node [right] {$\delta_1$}  (u1);
\draw[->] (dd1) -- node [above] {$\partial^*_0$} (dd2); \draw[->] (dd2) -- node [above] {$\partial^*_1$} (dd3); \draw[->] (dd3) -- node [above] {$\partial^*_2$} (dd4); ;
\draw[->] (dd2) --  node [right] {$\delta_0$} (d2); \draw[->] (dd3) --  node [right] {$\delta_0$} (d3); \draw[->] (dd1) --  node [right] {$\delta_0$} (d1);
\end{scope}\end{tikzpicture}\end{center}
where $\check{C}^k(\mathfrak{N}_{(l)},\tx{G})$ is the space of {\v C}ech $k$-cochains associated to the cover $\mathfrak{N}_{(l)}$ of $X_{(l)}$ and the horizontal coboundary maps are given by 
\begin{eqnarray*}
\partial_{k}^* = \sum_{i = 0}^{k+1} (-1)^{i} (\partial_{k+1}^i)^* 
\end{eqnarray*}
for $k \geq 1$ and $\partial^*_0 = t^* - s^*$. The maps $\delta_k$ are the standard {\v C}ech coboundary operators. 

It is inconvenient to describe geometric data in terms of cocycles in a double complex. Given a Lie groupoid $X_{\bullet}$ and a sheaf $\tx{G}$ as above one can also define the simplicial cohomology $H^*(X_{\bullet}, \tx{G})$ which is the cohomology of the complex
\begin{eqnarray*}
C^{\infty}(X_{(0)},G) \stackrel{\partial^*_0}{\longrightarrow} C^{\infty}(X_{(1)},G) \stackrel{\partial^*_1}{\longrightarrow} C^{\infty}(X_{(2)},G) \stackrel{\partial^*_2}{\longrightarrow} \cdots 
\end{eqnarray*}
In general, the simplicial cohomology is not isomorphic to the Lie groupoid cohomology. For example, any unit groupoid of a manifold has trivial cohomology in degrees greater than zero. In any case, there is a group homomorphism 
\begin{eqnarray}\label{i}
i: H^*(X_{\bullet}, \tx{G}) \rightarrow \check{H}^*(X_{\bullet}, \tx{G}_{\bullet})
\end{eqnarray}
which maps a cohomology class $[g] \in H^*(X_{\bullet}, \tx{G})$ to the bottom row in the double complex. If all the arrow manifolds $X_{(k)}$ are acyclic with respect to the {\v C}ech resolution then the {\v C}ech resolutions of the double complex become redundant and the group homomorphism $i$ is an isomorphism. In an ideal case, this can be arranged by defining a {\v C}ech groupoid as in 1.1 with respect to a suitable open cover of $X$. It is known that the groupoids arising from orbifolds have this property, \cite{MPro99}. 

We use the cohomology theory for classification of vector bundles and in the study of reduction and lifting problems of their structure groups. Therefore we need to consider sheafs of smooth functions valued in a nonabelian Lie group $G$. The degree one Lie groupoid cohomology groups $\check{H}^1(X_{\bullet}, \tx{G}_{\bullet})$ are well defined as usual, see e.g. \cite{TXL04}. When the degree one cohomology is considered, it is sufficient to apply a simple localization procedure by associating a {\v C}ech groupoid to any good open cover of the base manifold $X$. \3

\noindent \textbf{Lemma 1.} Let $\check{X}_{\bullet}$ be a {\v C}ech groupoid associated to a good open cover of $X$. Then the group homomorphism $i$ restricts to an isomorphism 
\begin{eqnarray*}
i: H^1(\check{X}_{\bullet}, \tx{G}) \rightarrow \check{H}^1(\check{X}_{\bullet}, \tx{G}_{\bullet}).
\end{eqnarray*}
 
\noindent Proof. The base $\check{X}_{(0)}$ in the {\v C}ech groupoid is a disjoint union of contractible sets and the leftmost {\v C}ech resolution in the double complex is acyclic. Any  representative $f$ of a cohomology class in $\check{H}^1(\check{X}_{\bullet}, \tx{G}_{\bullet})$ has components $f_{01} \in \check{C}^1(\mathfrak{N}_{(0)},\tx{G})$ and $f_{10} \in \check{C}^0(\mathfrak{N}_{(1)},\tx{G})$. By the acyclicity of the {\v C}ech resolution, we can find $h \in  \check{C}^0(\mathfrak{N}_{(0)},\tx{G})$ so that $d(h) + f = f'$ and $f'_{01} = 0$ ($d$ denotes the coboundary operator of the double complex). So any cohomology class in $\check{H}^1(\check{X}_{\bullet}, \tx{G}_{\bullet})$ can be represented by a class with a zero component in $\check{C}^1(\mathfrak{N}_{(0)},\tx{G})$. These are in the image of $i$ and so $i$ is surjective. 

For injectivity, suppose that there are $g_1$ and $g_2$ which represent classes of $H^1(\check{X}_{\bullet}, \tx{G})$ so that $[i(g_1)] = [i(g_2)]$ in $\check{H}^1(\check{X}_{\bullet}, \tx{G}_{\bullet})$. Thus, $i(g_1) = i(g_2) + d(h)$ for some $h \in \check{C}^0(\mathfrak{N}_{(0)},\tx{G})$. Because $i(g_1)$ and $i(g_2)$ have components only on the bottom row it follows that $\delta_0(h) = 0$ and so $h$ is a smooth function on $\check{X}_{(0)}$. Now, $g_1 = g_2 + \partial_0^*(h)$ and so $[g_1] = [g_2]$ in $ H^1(\check{X}_{\bullet}, \tx{G})$. Therefore $i$ is an isomorphism. \5 $\square$ \3

It is evident from the proof that the injectivity of the map \eqref{i} holds for all higher degree cohomology groups when $\tx{G}$ is determined by an abelian group. The surjectivity does not need to hold. 

Let $X_{\bullet}$ be a proper {\' e}tale groupoid and $\check{X}_{\bullet}$ a {\v C}ech groupoid associated to a good coordinate cover $\{N_a: a \in I\}$ of the base $X$. The tangent bundle $\tau X$ is subject to the action of the groupoid $\Theta$ as discussed in 1.3. This action determines the following structure cocycle 
\begin{eqnarray*}
g : \prod_{ab} \Theta_{N_b}^{N_a} \rightarrow GL_n, \5 g(\sigma) = (d \varphi_{\sigma})_{s(\sigma)}
\end{eqnarray*}
for all $\sigma \in \Theta_{N_b}^{N_a}$ where the differential is computed with respect to the coordinate charts in $N_b$ and $N_a$. When written out, the cocycle condition is just the compatibility of the action with the composition of arrows. Thus, $g$ represents a class in $H^1(\check{X}_{\bullet}, \tx{GL}_n)$. The components of $g$ arising from the unit arrows of $X_{\bullet}$ have a special importance. If $\textbf{1}_p \in \Theta_{N_{b}}^{N_{a}}$ this means that $p \in N_a \cap N_b \neq \emptyset$ and then 
\begin{eqnarray*}
g(\textbf{1}_p) = (d \iota)_p = J(p)_b^a                                              
\end{eqnarray*}
is just the usual Jacobian matrix computed from the change of coordinate charts.\3

\noindent \textbf{Proposition 2.} Let $X_{\bullet}$ be a riemannian proper {\' e}tale groupoid and $\check{X}_{\bullet}$ a {\v C}ech groupoid associated to a good coordinate cover $\{N_a\}$ of the base. Then the structure cocycle of $\tau X$ is valued in $O_n$ and if $\tau X$ is orientable the structure group has an $SO_n$ reduction. \3

\noindent Proof. The target group of the structure cocycle can be reduced to $O_n$ because the $\Theta$ action is orthogonal with respect to the riemannian structure. In the orientable case, the coordinate functions can be chosen so that the structure cocycle gets values in the group $GL_n^+$ and so the structure group has an $SO_n$ reduction. \5 $\square$ \3

In the case of a proper {\' e}tale action groupoid the orientability implies that the group acts by orientation-preserving isometries.

One can also reconstruct bundles from the cocycle data. In this process both, the geometric structure and the $\Theta$-action can be recovered. Moreover, only the cohomology class of the cocycle is important. \3

\noindent \textbf{Proposition 3.} Let $X_{\bullet}$ be the groupoid of Proposition 2. The tangent bundle $\tau X$ is fully determined by its structure cocycle $g$ in the simplicial cohomology group $H^1(\check{X}_{\bullet}, \tx{O}_n)$. Any element in the cohomology class of $g$ produces a vector bundle that is isomorphic to $\tau X$ as a vector bundle over $X_{\bullet}$. \3

An isomorphism of vector bundles over a Lie groupoid is an isomorphism of smooth vector bundles so that the isomorphism commutes with the $\Theta$-actions on both bundles. \3

\noindent Proof. A smooth vector bundle can be reconstructed from the cocycle data in the standard way. Define the total space of $\tau X$ by 
\begin{eqnarray*}
\tau X =  \Big[ \coprod_{a} N_a \times \mathbb{R}^n \Big] / \sim 
\end{eqnarray*} 
where $\sim$ is the equivalence relation which is determined by the restriction of the cocycle $g$ to the unit arrows of $X_{\bullet}$: if $\textbf{1}_p \in \Theta_{N_{b}}^{N_{a}}$  then $g(\textbf{1}_p) = (d \iota_p)$ and 
\begin{eqnarray*}
N_{b} \times \mathbb{R}^n \ni (p, v) \sim (p, (d \iota_p)_b^a v) \in N_{a} \times \mathbb{R}^n. 
\end{eqnarray*}
The trivialization $\vartheta_{N_a}$ over $N_a$ sends a pair $(p,v) \in N_a \times \mathbb{R}^k$ to its equivalence class $[p,v]$ in the bundle. The $\Theta$-action can be reconstructed by setting 
\begin{eqnarray*}
\sigma\cdot [p, v] = [\sigma p, (d \varphi_{\sigma}) v ],
\end{eqnarray*}
for any $\sigma \in \Theta_{N_{b}}^{N_{a}}$, $p \in N_{b}$ and $\sigma(p) \in N_{a}$. Now the unit arrows of $X_{\bullet}$ act as identities. 

Any cocycle that is cohomologous to $g$ is of the form 
\begin{eqnarray*}
&&g' : \prod_{ab} \Theta_{N_b}^{N_a} \rightarrow O_n \5 \text{with the local components} \\
&&g'(\sigma) = (f_{N_a} \circ t)(\sigma) g(\sigma) (f_{N_b} \circ s)^{-1}(\sigma) 
\end{eqnarray*}
for any $\sigma \in \Theta_{N_b}^{N_a}$ and some smooth functions $f_{N_b}: N_b \rightarrow O_n$ and $f_{N_a}: N_a \rightarrow O_n$. When restricted to the unit arrows of $X_{\bullet}$, the cocycle $g'$ defines a smooth vector bundle which is isomorphic to $\tau X$ and which is trivialized by 
\begin{eqnarray*}
\vartheta'_{N_a}(p,v) = \vartheta_{N_a}(p, f_{N_a}(p) v)  
\end{eqnarray*}
over $N_a$. This isomorphism respects the $\Theta$-action. Suppose that $\sigma \in \Theta_{N_b}^{N_a}$ is the arrow $p \mapsto \sigma p$, then 
\begin{eqnarray*}
\sigma \cdot \vartheta'_{N_b}(p,v) &=& \vartheta'_{N_b}(\sigma p,g'(\sigma)v) \\
&=& \vartheta_{N_b}(\sigma p, f_{N_a}(\sigma p) g(\sigma) v) \\
&=& \vartheta'_{N_a}(\sigma p, g(\sigma)v). 
\end{eqnarray*}
and so the bundle isomorphism which changes the trivialization commutes with the action. \5 $\square$ \3

\noindent \textbf{1.5.} Let $X_{\bullet}$ be a proper {\' e}tale groupoid and $\check{X}_{\bullet}$ a {\v C}ech groupoid associated to a good coordinate cover. Consider the exact group extension sequence  $1 \rightarrow SO_n \rightarrow O_n \rightarrow \mathbb{Z}_2 \rightarrow 0$ in which the second map is the inclusion $SO_n \hookrightarrow O_n$. This induces an exact sequence in the simplicial sheaf cohomology: 
\begin{eqnarray*}
  H^0(\check{X}_{\bullet}, \mathbb{Z}_2) \rightarrow H^1(\check{X}_{\bullet}, \tx{SO}_n) \rightarrow H^1(\check{X}_{\bullet}, \tx{O}_n) \rightarrow H^1(\check{X}_{\bullet}, \mathbb{Z}_2) 
\end{eqnarray*}
The image of the structure cocycle $g$ of $\tau X$ under the map in cohomology $q_*: H^1(\check{X}_{\bullet}, \tx{O}_n) \rightarrow H^1(\check{X}_{\bullet}, \mathbb{Z}_2)$ which is induced by the quotient map is called the first Stiefel-Whitney class of $X_{\bullet}$, 
\begin{eqnarray*}
q_*([g]) = w_1(X_{\bullet})  \in H^1(\check{X}_{\bullet}, \mathbb{Z}_2).
\end{eqnarray*}
The Stiefel-Whitney class is defined in the simplicial cohomology of the {\v C}ech groupoid $\check{X}_{\bullet}$. Recall that we can map this class to the Lie groupoid cohomology as in 1.4. \3

\noindent \textbf{Proposition 4.} The the orientability of $X_{\bullet}$ is equivalent to any of the following:
\begin{quote}
\textbf{1.} The class $w_1(X_{\bullet})$ represents the zero element in $H^1 (\check{X}_{\bullet}, \mathbb{Z}_2)$.
  
\textbf{2.} The class $i(w_1(X_{\bullet}))$ represents the zero element in $\check{H}^1 (\check{X}_{\bullet}, \mathbb{Z}_{2 \bullet})$.

\textbf{3.} There exists a nowhere vanishing and $\Theta$-invariant $n$-form on $X$.  
\end{quote}

\noindent Proof. The equivalence of the item 1 and the orientability follows from a standard diagram chase and proposition 3. The equivalence of 1 and 2 holds since $i$ is a group isomorphism. 

Suppose that $X_{\bullet}$ is orientable. Then $X$ is orientable as a manifold and there is a nowhere vanishing $n$-form $\Omega$. This $n$-form is $\Theta$-invariant since the determinants of $d \varphi_{\sigma}$ are equal to $1$ for all $\sigma \in \Theta$. So orientability implies 3. It remains to prove that 3 implies the orientability. Let $\Omega$ be a nowhere vanishing $\Theta$-invariant $n$-form. Fix the coordinate maps $\phi_a: N_a \rightarrow \mathbb{R}^n$ and denote by $\{x_i: 1 \leq i \leq n\}$ the standard euclidean coordinate system in $\mathbb{R}^n$. Then $\phi_a^*(dx_1 \wedge \cdots \wedge dx_n) = f_a \Omega$ for some nowhere vanishing function $f_a$ on $N_a$. We can assume that all $f_a$ are positive functions. Namely, if $f_c$ is negative on $N_c$, then we compose $\phi_c$ with an orientation reserving diffeomorphism $T: \mathbb{R}^n \rightarrow \mathbb{R}^n$ and replace the coordinate map with $\phi_c' = T \circ \phi_c$. For any $\sigma: x \rightarrow y$ in $\Theta_{N_b}^{N_a}$ there is a locally defined diffeomoprhism $\varphi_{\sigma}$ and the composition $\phi_a \circ \varphi_{\sigma} \circ \phi_b^{-1}$ is defined in a neighborhood of $\phi_b(x)$. Then using the $\Theta$-invariance we find that
\begin{eqnarray*}
& &(\phi_a \circ \varphi_{\sigma} \circ \phi_b^{-1})^* dx_1 \wedge \cdots \wedge dx_n \\
&=& (\varphi_{\sigma} \circ \phi_b^{-1})^* f_a \Omega \\
&=& (\phi_b^{-1})^* ((f_a \circ \varphi_{\sigma}) \Omega) \\
&=& (f_a \circ \varphi_{\sigma} \circ \phi_b^{-1}) (\phi_b^{-1})^* \Omega \\
&=&  \frac{f_a \circ \varphi_{\sigma} \circ \phi_b^{-1}}{f_b \circ \phi_b^{-1}} dx_1 \wedge \cdots \wedge dx_n. 
\end{eqnarray*}
Since all $f_a$ are positive functions, the transformations $d \varphi_{\sigma}$ have positive determinant. Since $\sigma$ is arbitrary, the orientability of $X_{\bullet}$ follows. \5 $\square$

\section{Spinor Bundles}

In this section we assume that $X_{\bullet}$ is an orientable proper {\' e}tale groupoid and $\check{X}_{\bullet}$ is a {\v C}ech groupoid constructed from a good coordinate cover $\{N_a\}$ of $X$. \3

\noindent \textbf{2.1.} Let  $\text{cl}(n)$ denote a real Clifford algebra generated by an $n$-dimensional vector space with a euclidean inner product. The spin group $\text{Spin}_n$ is a subgroup in the group of invertibles in $\text{cl}(n)$, see \cite{LM89}. There is a covering homomorphism $\varpi: \text{Spin}_n \rightarrow SO_n$. We have $\text{Spin}_1 = O_1$ and $\text{Spin}_2$ is isomorphic to $SO_2$. In the latter case $\varpi$ is a 2-fold covering homomorphism. For $n > 2$, $\varpi$ is a universal covering map with kernel equal to $\mathbb{Z}_2$. The complex spinor module $(\rho_s, \Sigma)$ is an irreducible complex representation for the complexification of $\text{cl}(n)$. The group $\text{Spin}_n$ acts on the spinor module through its embedding to the Clifford algebra and this gives the representation
\begin{eqnarray*}
\rho_s : \text{Spin}_n \rightarrow GL(\Sigma).
\end{eqnarray*}
The Clifford algebra is a $\text{Spin}_n$-module algebra under the adjoint action of $\text{Spin}_n$. More precisely, if $\gamma: \mathbb{R}^n \rightarrow \cl(n) \otimes \mathbb{C}$ is the canonical embedding, then for all $a \in \text{Spin}_n$ and $\gamma(u) \in \text{cl}(n) \otimes \mathbb{C}$ we get a group action 
\begin{eqnarray*}
\text{Ad}(a) \gamma(u) = a \gamma(u) a^{-1} = \gamma(\rho(\varpi(a))u)
\end{eqnarray*}
where $\rho: SO_n \rightarrow SO(\mathbb{R}^n)$ denotes the representation by matrix multiplication. 

The tangent spaces $\tau X_x$ equipped with the riemannian structure determine $n$-dimensional Clifford algebras at each $x \in X$. 

In the following we shall assume that $n \geq 2$. Consider the bundle $\tau X$ determined by the Lie groupoid cocycle $g$ defining a class in $H^1(\check{X}_{\bullet}, \tx{SO}_n)$. The cocycle $g$ can be pulled back through the covering morphism $\varpi: \text{Spin}_n \rightarrow SO_n$ which results a cochain 
\begin{eqnarray*}
\hat{g} := \varpi^*(g) : \prod_{ab} \Theta_{N_b}^{N_a} \rightarrow \text{Spin}_n. 
\end{eqnarray*}
This set of locally defined functions can be applied to define a bundle of Clifford algebras over $X_{\bullet}$. The fibres are the complexified Clifford algebras constructed from $\tau X_x$. Under the adjoint action of $\text{Spin}_n$ on the Clifford algebra, the center vanishes implying the relations 
\begin{eqnarray*}
\text{Ad}(\hat{g}(\tau)) \text{Ad}(\hat{g}(\sigma)) = \text{Ad}(\hat{g}(\tau   \sigma))
\end{eqnarray*}
for all composable pairs $(\tau, \sigma) \in \check{X}_{(2)}$. In fact, this makes $\text{Ad}(\hat{g})$ a cocycle in the cohomology group $H^1(\check{X}_{\bullet}, \tx{SO}(\cl(n)))$ where $\tx{SO}(\cl(n))$ denotes the sheaf of smooth functions getting values in the group of orientation preserving rotations on the vector space $\cl(n)$. Now the reconstruction determines a bundle of Clifford algebras which is trivialized over the cover $\{N_a\}$ and in which the $\Theta$-action is defined by 
\begin{eqnarray*}
\sigma \cdot [p, u_p] = [\sigma p , \text{Ad}(\varpi^*(d \varphi_{\sigma})_p)(u_p)].
\end{eqnarray*}
for $\sigma: p \rightarrow \sigma p$ and $u_p$ is an element in the fibre $\text{cl}_{\mathbb{C}}(\tau X_p)$. Let us denote by $\text{CL}(X_{\bullet})$ the Clifford bundle. 

In the space of sections the $\Theta$-action induces a pullback action by 
\begin{eqnarray*}
\varphi^{\#}_{\sigma}(e)_p =  \text{Ad}(\varpi^*(d \varphi_{\sigma})_p)^{-1}(e_{\sigma p }); \5 \sigma: p \rightarrow \sigma p.
\end{eqnarray*}
Given two sections of the Clifford bundle one can apply the Clifford multiplication fiberwise to define a product in the space of sections. Since the $\Theta$-action is the inverse adjoint action the multiplication and the $\Theta$-action commute with each other. This makes $\text{Sec}(\text{CL}(X_{\bullet}))$ a $\Theta$-module algebra. \3  

\noindent \textbf{2.2.} The cochain $\hat{g}$ is not necessarily a cocycle since the kernel of the covering morphism $\varpi$ is the subgroup $\mathbb{Z}_2$. It thus follows that although $g$ is a cocycle its lift through $\varpi$ does not need to be. The lifting problem is cohomological in nature. Associated to the lift there is the short exact sequence of groups
\begin{eqnarray*}
0 \rightarrow \mathbb{Z}_2 \rightarrow \text{Spin}_n \rightarrow SO_n \rightarrow 1
\end{eqnarray*}
This group extension sequence induces an exact sequence in the simplicial groupoid cohomology in the usual sense \cite{BX11}
\begin{eqnarray*}
 H^1(\check{X}_{\bullet}, \mathbb{Z}_2) \rightarrow H^1(\check{X}_{\bullet}, \tx{Spin}_n) \rightarrow H^1(\check{X}_{\bullet}, \tx{SO}_n) \rightarrow H^2(\check{X}_{\bullet}, \mathbb{Z}_2)
\end{eqnarray*}
The connecting homomorphism $\delta_*: H^1(\check{X}_{\bullet}, \tx{SO}_n) \rightarrow H^2(\check{X}_{\bullet}, \mathbb{Z}_2)$ is the map which applies the groupoid cohomology coboundary operator to the lifted cochain: $\delta_*(g) = \partial^*(\hat{g})$. Therefore the nontriviality of this class is an obstruction to lift $g$ to a groupoid cocycle and therefore an obstruction to define a spinor bundle with a compatible $\Theta$-action by the reconstruction from $\hat{g}$. Following the usual terminology we call 
\begin{eqnarray*}
\delta_*(g) := w_2(X_{\bullet}) \in H^2(\check{X}_{\bullet}, \mathbb{Z}_2)
\end{eqnarray*}
 the second Stiefel-Whitney class of the proper {\' e}tale groupoid $X_{\bullet}$. If this class is nontrivial, one gets a groupoid central extension which are classified by $H^2(\check{X}_{\bullet}, \tx{T})$, \cite{TXL04}. In this case, $\mathbb{Z}_2$ is viewed as a subgroup in $\mathbb{T}$. The proper {\' e}tale groupoid is called spin if $w_2(X_{\bullet})$ is trivial. The nontrivial cases give rise to groupoid spin gerbes, although they will not be discussed here.  \3

\noindent \textbf{Proposition 5.} Let $X_{\bullet}$ be an orientable proper {\' e}tale groupoid. The tangent bundle $\tau X$ can be lifted to the spinor bundle $F_{\Sigma}$ if and only if one of the following holds
\begin{quote}
\textbf{1.} The class $w_2(X_{\bullet})$ represents the zero element in $H^2(\check{X}_{\bullet}, \mathbb{Z}_2)$.

\textbf{2.} The class $i(w_2(X_{\bullet}))$ represents the zero element in $\check{H}^2(\check{X}_{\bullet}, \mathbb{Z}_{2 \bullet})$.
\end{quote}
Proof. The existence of the lift and 1 are equivalent by the discussion above. The equivalence of 1 and 2 holds since $i$ is an injective group homomorphism. \5 $\square$ \3

Finally, if $n = 1$ then $\cl(1) \otimes \mathbb{C} = \mathbb{C}$ and its irreducible complex representation is one dimensional. Since $X_{\bullet}$ is orientable, the structure cocycle of $\tau X$ gets values in $SO_1$ which is the trivial group. Therefore the spinor bundle in this case is just the trivial complex line bundle over $X$. Now the $\Theta$ action simply translates the fibres.

\section{Spectral Triple}

\noindent \textbf{3.1.} Consider a vector bundle $\xi$ over $X_{\bullet}$. Let $\Theta \fp{s}{\pi} \xi \rightarrow \xi$ be a left action of the arrows $\Theta$ on $\xi$. The action restricts to a linear isomorphism between the fibres and induces a smooth map $\Theta \fp{s}{\iota} X \rightarrow X$ on the base. For an arrow $\sigma \in \Theta_x$ we have a local bisection $\hat{\sigma}$ defined locally in a neighborhood $U_x$ of $x$ and the associated diffeomorphism $\varphi_{\sigma}$. Given a local section $\psi$ of $\xi$ defined in $\varphi_{\sigma}(U_x)$ we can pull it back to $U_x$ by 
\begin{eqnarray*}
\varphi^{\#}_{\sigma}(\psi)_p = \rho^{-1}(\hat{\sigma}(p)) (\psi \circ \varphi_{\sigma}(p))
\end{eqnarray*}
where $\rho$ is associated to the groupoid action, as in 1.3. This can be extended to the map 
\begin{eqnarray*}
&& \varphi^{\#}_{\sigma}: \Lambda^k(X, \xi)|_{\varphi_{\sigma}(U_x)} \rightarrow \Lambda^k(X,\xi)|_{U_x}; \\
&& (\varphi^{\#}_{\sigma}\Phi)(p, v_1, \ldots, v_k) = \rho^{-1}(\hat{\sigma}(p))\Phi(\varphi_{\sigma}(p), (d \varphi_p)v_1, \ldots, (d \varphi_p)v_k) 
\end{eqnarray*}
where $\Lambda^*(X, \xi)$ is the space of $\xi$-valued differential forms on $X$ and $v_i$ are tangent vectors at $p \in U_x$. 

A complex valued smooth function $f \in C^{\infty}(X)$ is $\Theta$-invariant if $f(x) = f(y)$ holds whenever there is an arrow $\sigma: x \rightarrow y$. This is equivalent to the invariance of $f$ under the local pullbacks $\varphi^{\#}_{\sigma}$  applied with the trivial representation $\rho$ on the complex line. Denote by $C^{\infty}(X)^{\Theta}$ the space of $\Theta$-invariant functions. A section $\psi$ in $\xi$ is called $\Theta$-invariant if it is invariant under the local pullbacks by the local diffeomorphisms $\Delta(\Theta)$.

Consider a vector bundle $\xi$ over $X_{\bullet}$. Then we call a connection $\nabla$ in $\xi$ a geometric connection if it is a connection in the differential geometric sense: $\nabla$ is a linear map $\nabla: \text{Sec}(\xi) \rightarrow \Lambda^1(X,\xi)$ which satisfies the Leibnitz rule
\begin{eqnarray*}
\nabla(f \psi) = d f \otimes \psi + f  \nabla(\psi)
\end{eqnarray*}
for all $f \in C^{\infty}(X)$ and $\psi \in \text{Sec}(\xi)$. A groupoid connection in $\xi$ is a geometric connection which is invariant under the pullbacks by the local diffeomorphisms $\Delta(\Theta)$:
\begin{eqnarray*}
\varphi_{\sigma}^{\#}  \nabla_{\sigma x}  (\varphi_{\sigma}^{\#})^{-1} = \nabla_x
\end{eqnarray*}
for all $\sigma \in \Theta_x$. 

Although geometric connections always exist and they form an affine space over the space of sections of the homomorphism bundle $\text{Hom}(\xi, \xi \otimes \tau X^*)$ it is not obvious that one can always find one with the $\Theta$-invariance property. In the case under consideration they do exist.\3

\noindent \textbf{Proposition 6.} If $\xi$ is a smooth vector bundle over a proper {\' e}tale groupoid $X_{\bullet}$. Then there is a groupoid connection.\3

\noindent Proof. Choose a cutoff $c: X \rightarrow \mathbb{R}_+$ and a Haar system in $X_{\bullet}$ (as defined in 1.2). Let $\nabla$ be any geometric connection in $\xi$. Then define 
\begin{eqnarray}\label{nablaI}
\nabla^I_x = \int_{\sigma \in \Theta_x} c(t(\sigma)) \varphi_{\sigma}^{\#}  \nabla_{\sigma x}  (\varphi_{\sigma}^{\#})^{-1} \mu_x(d \sigma). 
\end{eqnarray}
The function $c \circ t$ has a finite support over each $s$-fibre $\Theta_x$ and so the integration with respect to the Haar measure reduces to a finite sum fiberwise.
For each $x \in X$ and $\sigma \in \Theta_x$ the composition $\varphi_{\sigma}^{*}  \nabla_{\sigma x}  (\varphi_{\sigma}^{*})^{-1}$ maps a section of $\xi$ at $x \in X$ to a section of $\Lambda^1(X,\xi)$ at $x \in X$. Thus $\nabla^I_x$ is a map
\begin{eqnarray*}
\nabla^I_x : \text{Sec}(\xi)_x \rightarrow \Lambda^1(X,\xi)_x 
\end{eqnarray*}
Since the Haar integration is a smooth operation, the assignment $x \mapsto \nabla^I_x \psi_x$ defines a smoothly varying section in $\Lambda^1(X,\xi)$ for all $\psi \in \text{Sec}(\xi)$.

The linearity of $\nabla^I$ is obvious. For the Leibnitz rule we write locally $\nabla_{\sigma x} = d_{\sigma x} + A_{\sigma x}$ for all $\sigma \in \Theta_x$. Then 
\begin{eqnarray*}
\varphi_{\sigma}^{\#}  \nabla_{\sigma x}  (\varphi_{\sigma}^{\#})^{-1} &=& \rho(\hat{\sigma})^{-1} \varphi^*_{\sigma} (d_{\sigma x} + A_{\sigma x})  (\varphi^{-1}_{\sigma})^* \rho(\hat{\sigma}) \\
&=& d_x +  \rho(\hat{\sigma})^{-1}(d\rho(\hat{\sigma}))_x + \rho(\hat{\sigma})^{-1} \varphi^*_{\sigma} A_{\sigma x} (\varphi^{-1}_{\sigma})^* \rho(\hat{\sigma})\\
&=& d_x +  \rho(\hat{\sigma})^{-1}(d\rho(\hat{\sigma}))_x + \text{Ad}(\rho(\hat{\sigma}))^{-1} A_x .
\end{eqnarray*}
The exterior differential term is constant in the direction of the fibres of $s$ and therefore the connection $\nabla^I$ is of the form $d + A^I$ with $A^I \in \Lambda^1(X) \otimes \mathfrak{gl}_n$. A linear map $\text{Sec}(\xi) \rightarrow \Lambda(X,\xi)$ of this form satisfies the Leibnitz rule. 

For the invariance under $\Delta(\Theta)$, take $\tau \in \Theta$ be the arrow $x' \rightarrow x$. Then 
\begin{eqnarray*}
\varphi^{\#}_{\tau}  \nabla^I_{\tau x'} (\varphi^{\#}_{\tau})^{-1} &=& \int_{\sigma \in \Theta_x} c(t(\sigma)) \varphi^{\#}_{\tau}  \varphi^{\#}_{\sigma}  \nabla_{\sigma \tau x'}  (\varphi^{\#}_{\tau}  \varphi^{\#}_{\sigma})^{-1} \mu_x(d \sigma) \\
 &=&\int_{\sigma \in \Theta_x} c(t(\sigma \tau)) \varphi^{\#}_{\sigma \tau}  \nabla_{ \sigma \tau x'}  (\varphi^{\#}_{\sigma \tau})^{-1} \mu_x(d \sigma) \\
 &=& \int_{\sigma \in \Theta_{x'}} c(t(\sigma))  \varphi^{\#}_{\sigma}  \nabla_{\sigma x'}   (\varphi^{\#}_{\sigma})^{-1} \mu_{x'} (d \sigma) \\
&=& \nabla^I_{x'}. \5 \square
\end{eqnarray*}

Suppose that $\xi$ has an inner product so that $\Theta$ acts unitarily (orthogonally in the real case). The groupoid connection $\nabla$ is called hermitian (riemannian in the real case), if it is compatible with the inner product in the usual sense:  
\begin{eqnarray*}
d ( \psi_1,\psi_2 ) = ( \nabla \psi_1, \psi_2 ) + ( \psi_1, \nabla \psi_2 )
\end{eqnarray*}
for all smooth sections $\psi_1,\psi_2$ of $\xi$. \3
 
\noindent \textbf{Corollary 1.} If $\xi$ is a smooth vector bundle over a proper {\' e}tale groupoid $X_{\bullet}$. Then there is a hermitian (riemannian) groupoid connection. \3

\noindent Proof. Given any inner product in $\xi$ there is always a geometric connection which satisfies the compatibility condition. When written out in local coordinates, this means that the connection coefficients are $1$-forms with values in the Lie algebra $\mathfrak{u}_n$ ($\mathfrak{o}_n$ in the real case). The construction in the Proposition 6 now provides a hermitian (riemannian) groupoid connection. \5 $\square$ \3

\noindent \textbf{3.2.} Now we proceed towards the main goal. Suppose that $X_{\bullet}$ is an orientable spin proper {\' e}tale groupoid. Suppose that $\tau X$ is equipped with a fixed $\Theta$-invariant inner product. Associated to the inner product we have the Clifford bundle $\text{CL}(\tau X)$ and the spinor bundle $F_{\Sigma}$ over $X_{\bullet}$. The sections of the Clifford bundle act on the sections of the spinor bundle. Let us denote by $\text{Sec}(\text{CL}(\tau X))^{\Theta}$ and $\text{Sec}(F_{\Sigma})^{\Theta}$ the spaces of smooth sections that are invariant under the action of $\Delta(\Theta)$. Since the action on the former is by adjugation, it is obvious that the fiberwise Clifford multiplication determines a module structure: 
\begin{eqnarray*}
\text{Sec}(\text{CL}(\tau X))^{\Theta} \times \text{Sec}(F_{\Sigma})^{\Theta}  \rightarrow \text{Sec}(F_{\Sigma})^{\Theta}. 
\end{eqnarray*}
A groupoid connection in $\text{Sec}(\text{CL}(\tau X))$ clearly restricts to a linear map 
\begin{eqnarray*}
\nabla_{\text{CL}}: \text{Sec}(\text{CL}(\tau X))^{\Theta} \rightarrow \Lambda^1(X,\text{CL}(\tau X))^{\Theta}.
\end{eqnarray*}
The same holds for the groupoid connection $\nabla$ in $F_{\Sigma}$. If in addition $\nabla_{\text{CL}}$ and $\nabla$ are compatible with the module structure, then $\nabla$ is called Clifford compatible.\3

\noindent \textbf{Proposition 7.} If $X_{\bullet}$ is an orientable spin proper {\' e}tale groupoid, then there exists a Clifford compatible hermitian groupoid connection in the spinor bundle $F_{\Sigma}$.\3

\noindent Proof. We proceed by choosing geometric connections $\nabla_{\text{CL}}$ and $\nabla$ in $\text{CL}(\tau X)$ and in $F_{\Sigma}$ which are Clifford compatible and $\nabla$ is a hermitian connection. Then we follow the Proposition 6 and define the invariant connections $\nabla^I_{\text{CL}}$ and $\nabla^I$. It is hermitian by Corollary 1. For all $\sigma \in \Theta_x$, $e \in \text{Sec}(\text{CL}(\tau X))$ and $\psi \in \text{Sec}(F_{\Sigma})$ we have
\begin{eqnarray*}
\varphi_{\sigma}^{\#}  \nabla_{\sigma x} (\varphi_{\sigma}^{\#})^{-1} e \psi &=&
 \varphi_{\sigma}^{\#}  \nabla_{\sigma x} ((\text{Ad}(\varphi_{\sigma}^{\#})^{-1}  e)(( \varphi_{\sigma}^{\#})^{-1} \psi)) \\
&=& \varphi_{\sigma}^{\#} ((\nabla_{\text{CL}})_{\sigma x} (\text{Ad}(\varphi_{\sigma}^{\#})^{-1}  e)) (( \varphi_{\sigma}^{\#})^{-1} \psi)) \\ 
&+& \varphi_{\sigma}^{\#}((\text{Ad}(\varphi_{\sigma}^{\#})^{-1}  e) \nabla_{\sigma x} ( \varphi_{\sigma}^{\#})^{-1} \psi) \\
&=& (\text{Ad}(\varphi_{\sigma}^{\#}) (\nabla_{\text{CL}})_{\sigma x} \text{Ad}(\varphi_{\sigma}^{\#})^{-1}(e)) \psi_{x}  \\
&+&  e_x (\varphi_{\sigma}^{\#} \nabla_{\sigma x} (\varphi_{\sigma}^{\#})^{-1} \psi),
\end{eqnarray*}
where we have written Ad for the adjoint action of $\Delta(\Theta)$ on the Clifford sections. Then
\begin{eqnarray*}
\nabla^I(e\psi) = (\nabla_{\text{CL}}^I e) \psi + e \nabla^I \psi
\end{eqnarray*}
follows from the formula \eqref{nablaI}.  \5 $\square$ \3

\noindent \textbf{3.3.} Let us fix a groupoid riemannian structure in the bundle $\tau X$ over $X_{\bullet}$. Locally we can choose $n$ linearly independent vector fields $e_i$ of $\tau X$ with the dual vector fields $e_i^*$ with respect to the riemannian structure. Associated to the riemannian structure we have Clifford and spinor bundles over the groupoid. Let $\nabla$ be a Clifford compatible hermitian groupoid connection in $F_{\Sigma}$. The Dirac operator is a differential operator $\eth: \text{Sec}(F_{\Sigma}) \rightarrow \text{Sec}(F_{\Sigma})$ given by 
\begin{eqnarray*}
\eth = \sum_{i = 1}^n \gamma(e_i^*)  \nabla_{e_i}. 
\end{eqnarray*}
From the point of view of the base manifold $X$ (as an ordinary manifold), $\eth$ is an ordinary Dirac operator acting on the space of smooth spinor fields as usual. Fundamental is the following.\3

\noindent \textbf{Proposition 8.} The Dirac operator is invariant under the $\Theta$-action. In particular, it can be restricted to $\text{Sec}(F_{\Sigma})^{\Theta}$. \3

\noindent Proof. If $u$ is a vector field in $\tau X$, then the invariance of $\nabla$ implies
\begin{eqnarray*}
\varphi^{\#}_{\sigma} \nabla_{u_{\sigma x}} (\varphi^{\#}_{\sigma})^{-1} =  \nabla_{(d \varphi_{\sigma})^{-1}  u_x} .  
\end{eqnarray*}
for any $\sigma \in \Theta_x$. It follows 
\begin{eqnarray*}
\varphi^{\#}_{\sigma} \eth_{ \sigma x} (\varphi^{\#}_{\sigma})^{-1} &=& \sum_{i = 1}^n (\text{Ad}(\varphi^{\#}_{\sigma})(\gamma (e_i^*)))_{\sigma x}  \varphi^{\#}_{\sigma} \nabla_{(e_i)_{\sigma x}} (\varphi_{\sigma}^{\#})^{-1} \\
&=& \sum_{i = 1}^n \text{Ad}(\rho(\hat{\sigma})^{-1}) \gamma (e_i^*)_x  \nabla_{(d \varphi_{\sigma})^{-1} (e_i)_x} \\
&=& \sum_{i = 1}^n \gamma (((d \varphi_{\sigma})^{-1}) e_i^*)_x  \nabla_{(d \varphi_{\sigma})^{-1} (e_i)_x}. \\
&=& \eth_x.
\end{eqnarray*}
The last step holds since the element $\sum_i (e_i^*)_x \otimes (e_i)_x$ transforms as a trivial representation under the action of $O_n$ for all $x \in X$. \5 $\square$ \3

To make $\eth$ a formally self-adjoint operator we need to set one more condition for the Clifford structure. Namely, the unit vector fields $\gamma(u)$ in $\text{Sec}(\text{CL}(\tau X))$ need to act on the spinors unitarily 
\begin{eqnarray*}
(\gamma(u) \psi_1 , \gamma(u) \psi_2) = (\psi_1, \psi_2). 
\end{eqnarray*}
A $\Theta$-invariant inner product which satisfies this conditions exists. To see this it is sufficient to choose an inner product with this property and then average out the $\Theta$ action as in section 1.2. A complex spinor bundle is called a complex Dirac bundle on $X_{\bullet}$ if it is equipped with a hermitian $\Theta$-invariant groupoid connection which is Clifford compatible, and a $\Theta$-invariant inner product which is normalized so that the units of $\text{Sec}(\text{CL}(\tau X))$ act as unitary transformations. 

Let $\text{Sec}(F_{\Sigma})$ denote the space of smooth sections in the spinor bundle and let $(\cdot, \cdot) : \text{Sec}(F_{\Sigma}) \times \text{Sec}(F_{\Sigma}) \rightarrow C^{\infty}(X)$ be the natural pairing in the fibres. \3 

\noindent \textbf{Proposition 9.} A complex Dirac bundle $F_{\Sigma}$ exists on an orientable spin proper {\' e}tale groupoid. The Dirac operator satisfies
\begin{eqnarray}\label{prop2}
(\psi_1, \eth \psi_2) - (\eth \psi_1, \psi_2) = \text{div}(V) 
\end{eqnarray}
where $\psi_i \in \text{Sec}(F_{\Sigma})^{\Theta}$ and $V$ is a vector field determined by the condition that $(V,W) = - (\psi_1, \gamma(W) \psi_2)$ for all vector fields $W$ on $X$. \3

\noindent Proof. The existence of a complex Dirac bundle is proved above. The second part can be proved as in the usual geometric case, \cite{LM89} II.5.3. \3

Notice that the divergence on the right side of \eqref{prop2} is $\Theta$ invariant (because the left side is). \3

\noindent \textbf{3.4.} The Dirac operator $\eth$ fits in the definition of a spectral triple only if we complete the space of invariant spinor sections in such a way that $\eth$ is a formally self-adjoint. We exploit the orbifold structure in the orbit space $|\Theta|$ to define an integration of $\Theta$-invariant functions. We restrict to the compact case: $X_{\bullet}$ is defined to be compact if the orbit space $|\Theta|$ is compact. Compactness is independent on the Morita equivalence class since the Morita equivalence leaves $|\Theta|$ invariant. 
 
Suppose that $X_{\bullet}$ is compact. If $x \in X$, then the groupoid $X_{\bullet}$ localizes to a transformation groupoid $G \ltimes U \rightrightarrows U$ where $U$ is a connected open neighborhood of $x \in X$. Concretely we have a smooth action of a finite group on the manifold $U$ and the orbit space $|U| = U/G$ has a stratification by orbit types. Let $U^0$ denote the preimage of the principal stratum of $|U|$ under the projection. Let $k$ be the order of the isotropy group $G_x$ at some point $x \in U^0$. Since the isotropy is constant along the strata, the choice of $x \in U^0$ is arbitrary. Suppose that $X_{\bullet}$ is a riemannian groupoid. Then the invariant riemannian volume form $\nu$ on $X_{\bullet}$ restricts to a $G$-invariant volume form on $U$ and the integration of a $G$-invariant function on $U$ is defined by 
\begin{eqnarray*}
\int_{|U|} f = \frac{k}{|G|}\int_{U} f \nu. 
\end{eqnarray*}
where $|G|$ denotes the rank of $G$ (the rank is always finite since the groupoid $\Theta$ is proper and {\' e}tale).

Suppose that the cover of $X$ is chosen so that $X_{\bullet}$ localizes to an action groupoid $\Theta_a^a \ltimes N_a \rightrightarrows N_a$ over each $N_a$ so that $\Theta_a^a$ is an isotropy group at some $x \in N_a$, recall 1.1. The image of $N_a$ under the groupoid projection, $|N_a|$, is an open subset of $|\Theta|$. The subsets $|N_a|$ define an open cover of $|\Theta|$. It is well known in the orbifold theory that there is a partition of unity: a collection of functions $\{\varrho_a: N_a \rightarrow \mathbb{R}\}$ such that $\varrho_a$ is $\Theta_a^a$-invariant and
\begin{quote}
\textbf{1.} $0 \leq \varrho_a \leq 1$, 

\textbf{2.} $\text{supp}(\varrho_a) \subset N_a$,

\textbf{3.} $\sum_{a} |\varrho_a|(p) = 1$,  for all $p \in |\Theta|$. 
\end{quote}
The function $|\varrho_a|$ in the condition 3 is the continuous function on $|N_a|$ that is covered by the $\Theta_a^a$-invariant function $\varrho_a$ on $N_a$, i.e. the value of $|\varrho_a|$ at $p$ is equal to the value of $\varrho_a$ at any point in the orbit that is represented by $p$. Let $f \in C^{\infty}(X)^{\Theta}$. The integration of $f$ is defined by 
\begin{eqnarray*}
\int_{|\Theta|} f = \sum_{a} \frac{k_a}{|\Theta_a^a|}\int_{N_a} \varrho_a f \nu
\end{eqnarray*}
where $k_a$ is the rank of the isotropy along the preimage of the principal stratum in $N_a$. Define an inner product in $\text{Sec}(F_{\Sigma})^{\Theta}$ by 
\begin{eqnarray}\label{innerP}
\langle \psi_1, \psi_2 \rangle = \int_{|\Theta|} (\psi_1, \psi_2). 
\end{eqnarray}
The integration over a compact orbifold structure satisfies the stokes theorem. So Proposition 9 gives: \3

\noindent \textbf{Proposition 10.} If $X_{\bullet}$ is compact spin proper {\' e}tale groupoid, then the Dirac operator acting on the space of invariant sections $\text{Sec}(F_{\Sigma})^{\Theta}$ is formally self-adjoint. \3

\noindent \textbf{3.5.}  A spectral triple for the algebra $C^{\infty}(X)^{\Theta}$ is data $(C^{\infty}(X)^{\Theta}, \hil, \eth)$ where $\hil$ is a complex Hilbert space on which $C^{\infty}(X)^{\Theta}$ has a faithful bounded $*$-representation and the Dirac operator $\eth$ acts as a densely defined self-adjoint operator such that $[\eth, f]$ is a bounded operator for all $f \in C^{\infty}(X)^{\Theta}$. Whenever $X$ is even dimensional the spectral triple is required to have a chiral grading, namely a bounded operator $\omega$ with 
\begin{eqnarray}\label{gamma}
\omega^2 = 1,\5 \{\omega, \eth\} = [\omega, f] = 0
\end{eqnarray}
for all $f \in C^{\infty}(X)^{\Theta}$. Moreover, one says that the spectral triple is finitely summable if $(1 + \eth^2)^{-m}$ is a trace class operator for some finite $m \in \mathbb{N}$. In the odd dimensional case $\omega$ is defined to be the unit operator, $\omega = 1$. 

The complex Hilbert space can be constructed by a completion of the space of $\Theta$-invariant sections $\text{Sec}(F_{\Sigma})^{\Theta}$ with respect to the $L^2$-inner product \eqref{innerP}. Let us denote by $L^2(F_{\Sigma})^{\Theta}$ this completion. Suppose that the Dirac operator $\eth$ has an extension on $L^2(F_{\Sigma})^{\Theta}$ which is still denoted by $\eth$. The Clifford module has a canonical section defined by  
\begin{eqnarray*}
\omega = i^{\frac{n}{2}} \gamma(e_1) \cdots \gamma(e_n)
\end{eqnarray*}
if $e_i: i \in \{1, \ldots, n\}$ define an orthonormal basis fiberwise. The section $\omega$ transforms as a volume form, and since the adjoint action of $\Theta$ acts on the vector fields as orientation preserving rotations ($X_{\bullet}$ is orientable), $\omega$ is $\Theta$-invariant. In particular, $\omega$ acts on $\text{Sec}(F_{\Sigma})^{\Theta}$ as a bounded operator and so $\omega$ extends to a bounded operator on $L^2(F_{\Sigma})^{\Theta}$. This operator satisfies \eqref{gamma} if $n$ (the dimension of $X$ and $\Theta$) is even. If $n$ is odd, then $\omega$ acts as the unit operator. \3

\noindent \textbf{Theorem 1.} Let $X_{\bullet}$ be a compact orientable spin proper {\' e}tale groupoid and let $F_{\Sigma}$ be a Dirac bundle on $X_{\bullet}$. Then  
\begin{eqnarray*}
(C^{\infty}(X)^{\Theta}, \eth, L^2(F_{\Sigma})^{\Theta})
\end{eqnarray*}
defines a finitely summable spectral triple, and if $X$ is even dimensional, then the spectral triple is equipped with a chirality operator $\omega$.\3

\noindent Proof. It is obvious that the algebra $C^{\infty}(X)^{\Theta}$ has an action on $\text{Sec}(F_{\Sigma})^{\Theta}$ by pointwise multiplication. This extends to a  bounded $*$-representation on $L^2(F_{\Sigma})^{\Theta}$. The $*$-structure is clearly $\Theta$-invariant and the representation is faithful.  The standard proof of self-adjointness of the closure of the Dirac operator (\cite{GVF01} Theorem 9.15) can be applied in this setup straightforwardly: the proof only requires self-adjointness in a dense domain, which in our case is $\text{Sec}(F_{\Sigma})^{\Theta}$, existence of a finite partition of unity and the property that $\eth$ is a pseudodifferential operator of order $1$ on $X$, which holds since $\eth$ is a classical Dirac operator. Similarly we see that $[\eth, f]$ has to be bounded on $L^2(F_{\Sigma})^{\Theta}$ since the commutators $[\eth, f]$ are given  exactly as in the classical case, \cite{LM89} II.5.5. The finite summability of $\eth$ holds since $\text{Sec}(F_{\Sigma})^{\Theta}$ is a subspace in the space of all sections of the spinor bundle $F_{\Sigma}$ and the Dirac spectrum satisfies the finite summability in the latter case since $X$ is a finite dimensional manifold. \5 $\square$ \3

\noindent \textbf{3.6.} The orientability of the spectral triple on $C^{\infty}(X)^{\Theta}$ is equivalent to the existence of a Hochischild cycle
\begin{eqnarray*}
c = \sum_{i = 1}^k a^0_i \otimes a^1_i \otimes \cdots \otimes a^n_i  \in Z_n(C^{\infty}(X)^{\Theta},C^{\infty}(X)^{\Theta})
\end{eqnarray*}
such that when represented on $L^2(F_{\Sigma})^{\Theta}$
\begin{eqnarray}\label{orientation}
\sum_{\i = 1}^k a^0_i[\eth, a^1_i] \cdots [\eth, a^n_i] = \omega,
\end{eqnarray}
where $\omega$ is the chirality operator \eqref{gamma}. It is proved in \cite{RV08} that in the case of a global action orbifold, the orientation cannot be satisfied unless the group action on $X$ is free. This is because of the lack of invariant functions to construct the Hochschild orientation cycle. The orientation holds in the following weaker form: there exists a Hochschild cycle
\begin{eqnarray*}
c = \sum_{i = 1}^k a^0_i \otimes a^1_i \otimes \cdots \otimes a^n_i  \in  Z_n(C^{\infty}(X),C^{\infty}(X))
\end{eqnarray*}
such that \eqref{orientation} holds and $\omega$ is the chiral grading operator on the $\Theta$-invariant Hilbert space $L^2(F_{\Sigma})^{\Theta}$. According to 3.5, the weakened orientation holds in the case of the Dirac spectral triple. 

If $X$ is a compact manifold, one can identify the differential $n$-forms with the Hochschild $n$-homology classes of the algebra $C^{\infty}(X)$. The reason why the usual spectral triple orientation fails in the study of invariant smooth functions is that, unless the groupoid acts freely, one cannot identify the invariant $n$-forms with the Hochschild $n$-homology classes of $C^{\infty}(X)^{\Theta}$. In particular, according to Proposition 4, proper {\' e}tale groupoids have a nowhere vanishing and invariant orientation form
\begin{eqnarray*}
\nu = \sum_{\i = 1}^k a^0_i da^1_i \wedge \cdots \wedge da^n_i
\end{eqnarray*}
where $a_i^k \in C^{\infty}(X)$ but according to \cite{RV08} one cannot have $a_i^k \in C^{\infty}(X)^{\Theta}$ for all the indices $i$ and $k$. \3

\noindent \textbf{3.7.} It is evident from the discussion above that the algebra of smooth invariant functions is not large enough to provide a good spectral triple. Therefore one needs an extension of the invariant spectral triple. Another algebra of interest is the algebra of smooth compactly supported complex valued functions on $\Theta$ equipped with the convolution product. Denote by $C^{\infty}_c(\Theta)$ this algebra. A spectral triple associated with $C^{\infty}_c(\Theta)$ was studied in \cite{GL13} in the context of  effective proper {\' e}tale Lie groupoids and Dirac type operators. Here we take $X_{\bullet}$ to be an effective orientable spin proper {\' e}tale groupoid and prove that in this special case, the construction detailed in this work provides an explicit model for a spectral triple on $C^{\infty}_c(\Theta)$. Let $L^2(F_{\Sigma})$ be the space of square integrable sections of the $\Theta$-equivariant Dirac bundle. The integration on $X$ is applied in the definition of the $L^2$-inner product: recall that the groupoid orientability implies the orientability of its base manifold, 1.3. The algebra $C^{\infty}_c(\Theta)$ can be represented on $L^2(F_{\Sigma})$ by
\begin{eqnarray}\label{convolution}
(f \cdot \psi)_x = \int_{\sigma \in \Theta^x} f(\sigma) (\varphi^{\#}_{\sigma^{-1}} \psi)_{x} \mu^x (d \sigma)
\end{eqnarray}
for all $f \in C^{\infty}_c(\Theta)$ and $\psi \in L^2(F_{\Sigma})$. The measures $\mu^x$ are left invariant measures in the fibres $\Theta^x$ of the target maps and vary smoothly on $X$. The Haar integration is well defined over each $\Theta^x$ because the function $f$ is compactly supported in these fibres. \3

\noindent \textbf{Theorem 2.} Let $X_{\bullet}$ be an effective orientable spin proper {\' e}tale groupoid and let $F_{\Sigma}$ be a Dirac bundle on $X_{\bullet}$. If the base manifold $X$ is complete, then    
\begin{eqnarray*}
(C^{\infty}_c(\Theta), \eth, L^2(F_{\Sigma}))
\end{eqnarray*}
defines a finitely summable spectral triple. \3

\noindent Proof. Since $X$ is complete, $\eth$ has a  self-adjoint closure on $L^2(F_{\Sigma})$ which makes $\eth$ finitely summable. The action \eqref{convolution} defines a faithful $*$-representation which preserves the domain of $\eth$, \cite{GL13}. 

It remains to verify that the commutators $[\eth, f]$ extend to bounded operators on $L^2(F_{\Sigma})$. For each arrow $\sigma$ the target map $t$ has a local smooth inverse in a neighborhood of $t(\sigma)$ which we denote by $\widetilde{\sigma}$.  The discrete measures $\mu^x$ vary smoothly over $X$. If $\psi \in \text{Sec}_c(F_{\Sigma})$ and $f \in C^{\infty}_c(\Theta)$, then using the $\Theta$-invariance of the Dirac operator we get
\begin{eqnarray*}
 [\eth,f]_x  \psi_x &=&  \eth_x \int_{\sigma \in \Theta^x} (f \circ \widetilde{\sigma})_x (\varphi^{\#}_{\sigma^{-1}} \psi)_{x} \mu^x(d \sigma) - (f \cdot \eth)_x \psi_x\\
&=& \sum_i \gamma(e_i^*)_x \int_{\sigma \in \Theta^x} e_i (f \circ \widetilde{\sigma})_x (\varphi^{\#}_{\sigma^{-1}} \psi)_{x} \mu^x(d \sigma) \\
&+&  \int_{\sigma \in \Theta^x} f(\sigma) (\varphi^{\#}_{\sigma^{-1}}(  \eth_{\sigma^{-1} x} \psi))_{x}  \mu^x(d \sigma) \\
&+& \sum_i \gamma(e_i^*)_x \int_{\sigma \in \Theta^x} f(\sigma) (\varphi^{\#}_{\sigma^{-1}} \psi)_{x} e_i (\mu)^x(d \sigma)  - (f \cdot \eth)_x \psi_x \\
&=&\sum_i \gamma(e_i^*)_x \int_{\sigma \in \Theta^x} e_i (f \circ \widetilde{\sigma})_x (\varphi^{\#}_{\sigma^{-1}} \psi)_{x} \mu^x(d \sigma)  \\
&+& \sum_i \gamma(e_i^*)_x \int_{\sigma \in \Theta^x} f(\sigma) (\varphi^{\#}_{\sigma^{-1}} \psi)_{x}  e_i (\mu)^x(d \sigma).
\end{eqnarray*}
Since the function $f$ and the measure $\mu$ are smooth in $X$ we see that $[\eth,f]$ extends to a bounded operator on $L^2(F_{\Sigma})$. \5 $\square$ \3

Suppose that $X_{\bullet}$ is the groupoid of Theorem 2 and that $C_c^{\infty}(\Theta)$ is a unital algebra. Denote by $\textbf{1}$ the unit for the convolution product. If $g \in C^{\infty}(X)$, then 
\begin{eqnarray*}
((g \circ t) \textbf{1} \cdot \psi)_x =  g \psi_x.
\end{eqnarray*}
In particular, we find a Hochschild cycle in $Z_n(C_c^{\infty}(\Theta),C_c^{\infty}(\Theta))$ which gives rise to the $\Theta$-invariant chirality section of $\text{CL}(X_{\bullet})$ under the assignment \eqref{orientation}. This section extends to a bounded operator $\omega$  on $L^2(F_{\Sigma})$ and one can check using the $\Theta$-invariance of $\omega$ that the  operator commutators $[\omega, f]$ are zero for all $f \in C_c^{\infty}(\Theta)$. So, the spectral triple of Theorem 2 satisfies the standard orientability condition.

\bibliographystyle{plain}

\end{document}